\documentclass[aps,prl,twocolumn,showpacs]{revtex4}
\usepackage{graphicx}
\usepackage{epsf}
\usepackage{psfrag}
\usepackage[T1]{fontenc} %Esconlha do pacote de fontes a ser utilizado!
\usepackage{epsfig,latexsym}
\usepackage{color}
\usepackage[latin1]{inputenc}
\usepackage{amsmath}
\usepackage{amssymb}
\usepackage{amsfonts}
\usepackage{indentfirst}
\usepackage{ae}
\usepackage{float}

\newcommand{\bq}{\begin{eqnarray}}
\newcommand{\eq}{\end{eqnarray}}

\begin{document}

\title{Nonlocal single particle correlations and EPR state in the double-slit experiment}
\author{O. R. de Araujo$^{1}$, H. Alexander$^{2}$, Marcos Sampaio$^{2}$, and I. G. da Paz$^{1}$}

\affiliation{$^1$ Departamento de F\'{\i}sica, Universidade Federal
do Piau\'{\i}, Campus Ministro Petr\^{o}nio Portela, CEP 64049-550,
Teresina, PI, Brazil}

\affiliation{$^{2}$ Departamento de F\'{\i}sica, Instituto de
Ci\^{e}ncias Exatas, Universidade Federal de Minas Gerais, Caixa
Postal 702, CEP 30161-970, Belo Horizonte, Minas Gerais, Brazil}

\begin{abstract}
Quantum correlations of observables for two particle states have
demonstrated the nonlocal character of the quantum mechanics.
However nonlocality can be exhibited even for noncommuting
observables of a single particle system. In this paper we show
nonlocality of  position-momentum correlations of a single particle
in the double-slit experiment modeled by an initially correlated
Gaussian wavepacket. The positivity or negativity of the Wigner
function for the state of the particle at the detection screen is
related with the $\sigma_{xp}$ covariances. A Bell's inequality is
constructed from the Wigner function and it is violated for both the
positive and negative cases. The case of positive Wigner function is
the analogous of the original EPR state for a single particle.
\end{abstract}


\maketitle

\section{Introduction}

Entanglement and nonlocality has been extensively studied since
Einstein, Podolsky and Rosen (EPR) put forward a gedanken experiment
in 1935 questioning the completeness of quantum mechanics \cite{EPR}
in which two particles were entangled simultaneously over a
continuum of position and momentum states. The crucial point in EPR
reasoning was that the position and momentum of the unmeasured
particle were simultaneous realities and thus violated Heisenberg's
uncertainty relation. Nonlocal correlations involving discrete
variables were cast later on by Bohm in 1951 \cite{Bohm}.The
interest in studying nonlocal quantum correlations has grown because
they are fundamental resources in field of quantum information
science \cite{Horodecki}. The Bell's inequality \cite{Bell1},
derived in 1964, served to prove right EPR's disturbing action at a
distance. It has become an important tool to investigate nonlocality
effects in discrete variable systems such as two entangled photons
in a cascade experiments \cite{Clauser} and in  parametric
down-conversion \cite{Kwiat1}, as well as entangled states of
trapped ions \cite{Rowe}. Bell's inequality violation was also
studied for a single particle by considering single photons
entangled in momentum and polarization \cite{Gadway}. Recently a
loophole free experiment showed the Bell's inequality violation
using electron spins separated by $1.3$ kilometres \cite{Hensen},
after the seminal experiment of Aspect and collaborators
\cite{Aspect1}.
%%%%%%%%%%%%%%%%%%%%

Of particular interest in quantum information tasks is a
demonstration of nonlocality in systems described by continuous
variables \cite{Franson}-\cite{Gatti} such as the original EPR state
\cite{Bell2} or the two-mode squeezed state \cite{Pati} with the aid
of dichotomic observables e.g. pseudo-spins \cite{Chen}, and parity
observables \cite{Saleh}. In \cite{Reid} it was developed an EPR
criterion which could be implemented with momentum and position-like
quadrature observables of squeezed states of light. Moreover, in
\cite{Howell} a demonstration of the EPR paradox using position- and
momentum-entangled photon pairs produced by spontaneous parametric
downconversion. They found that the position and momentum
correlations allow the position or momentum of a photon to be
inferred from that of its partner with a product of variances $\le
0.01 \hbar^2$, violating the separability bound by two orders of
magnitude. The EPR paradox does not represent a true inconsistency
because as the measurement involves only one quantity or the other,
the position and momentum of the unmeasured particle need not be
simultaneous realities.

Although Bell-type experiments involve multiple particles, the
non-commuting nature should be independent of whether the system
consists of multiple particles or a single particle. Entanglement
between two degrees of freedom has been demonstrated in single
neutron interferometry experiments \cite{Hasegawa}. This is
interesting because a violation of a Bell inequality would serve to
verify the uncertainty principle since it would indicate a definite
correlation between, say, position and momentum \cite{Durham}.

On the other hand quantum correlations
of continuous variables can be analyzed in the phase-space using the
Wigner function. The latter is an important tool to study
nonlocality in continuous-variable systems. Moreover, the Wigner
function can be measured in different system configurations
\cite{Banaszek0} and calculated for
arbitrary quantum systems \cite{Tilma}. Previously, there have been
efforts in understanding the relation between Bell nonlocality and
the Wigner function \cite{Cetto,Leonhardt,Johansen,Cohen}. Such
studies show that the positive definite Wigner function of the EPR
state can be used with a natural phase-space framework in which the
nonlocal character of this state can be studied \cite{Banaszek}.

In the simple Gaussian minimum-uncertainty wavepacket solution for
the Schr\"{o}dinger equation for a free particle,  $\sigma_{xp}$
(hereafter also called $\sigma_{xp}$ correlations) at $t=0$ are zero
but develop at later times as a result of the quantum dynamics
\cite{Bohm,Saxon}. However, more complex states such as squeezed
states or linear combination of Gaussian states can exhibit
initially such correlations \cite{Robinett, Riahi,Dodonov,Campos}.
Such $\sigma_{xp}$ correlations can be related with  phases of the
wave function \cite{Bohm}. They were also shown to play a role in
matter wave slit diffration: for example, qualitative changes in the
interference pattern appear as $\sigma_{xp}$ correlations develop
\cite{Carol}. More specifically, the Gouy phase of matter waves is
directly related with these correlations, as studied by the first
time in Refs. \cite{Paz1}. More recently, it was shown that the
maximum of these correlations is related with the minimum number of
interference fringes in the double-slit experiment \cite{solano}.
Double-slit experiments have been used to elucidate fundamental
aspects of the quantum theory \cite{Feynman}. The wave-particle
duality has been observed in the double-slit experiment with
electrons \cite{Jonsson}, neutrons \cite{Zeilinger1} and atoms
\cite{Carnal}. Recently, a controlled electron double-slit
diffraction was experimentally realized in which the probability
distributions for single- and double slit arrangements were observed
\cite{Bach}.

In this work, we use the double-slit setup to study nonlocality
associated with noncommuting observables for a single particle. This
setup enable us to connect the $\sigma_{xp}$ correlations with the
positivity or negativity of the Wigner function. The nonlocal
character of correlations for the position and momentum  of a single
particle is obtained by constructing a Bell's inequality using the
Wigner function.  For the sake of  calculability we consider an
initial Gaussian wavepacket and Gaussian shaped slit apertures. The
single particle state at the detection screen will be a
superposition of two Gaussian. Hence, after the slits, the single
particle state is represented by two parts, which is essential to
observe single particle nonlocal correlations analogous to a two
particle EPR system.

The behavior of $\sigma_{xp}$ correlations in an initially
$xp$-correlated Gaussian wavepacket measured by a real parameter
$\rho$ yield crucial information about the Wigner function. Whilst
for $\rho\geq 0$ the correlation is maximum for a given propagation
time (and  maximum correlation in turn  leads to maximum negativity
of the Wigner function)  for $\rho<0$ these correlations have a
maximum {\it{and}} a minimum for specific propagation times. As we
shall see the minima of $\sigma_{xp}$ correlations will be related
with a positive Wigner function. The state at the detection screen
with maximum (minimum)  $\sigma_{xp}$ correlations gives the maximum
(minimum) value for the Bell inequality violation. Moreover the
state at the detection screen (superposition of two Gaussian waves)
with minimum $\sigma_{xp}$ correlations has a positive definite
Gaussian shaped Wigner function. A positive Gaussian shaped Wigner
function has been used to study the original EPR state
\cite{Banaszek}. As we shall see, the state at the detection screen
with minimum $\sigma_{xp}$ correlations can be considered as
analogous of the original EPR state for a single particle.

In section II we model the double-slit experiment with matter waves
considering an initially correlated Gaussian wavepacket. The initial
wavepacket propagates during the time $t$ from the source to the
double-slit and during the time $\tau$ from the double-slit to the
screen. We calculate the wave functions after the passage through each
slit using the Green's function for the free particle. In section
III, we calculate the Wigner function and the $\sigma_{xp}$
correlations for the state that is a linear combination of the
states which passed through each slit. We show that the minimum
$\sigma_{xp}$ correlations are related with a positive definite
Wigner function and the maximum $\sigma_{xp}$ correlations with a
Wigner function with negative part. In section IV, we construct a
Bell-type inequality for the position and momentum observables using
the Wigner function and show the nonlocal character for a single
particle. The condition for the production of an analogous EPR state
for a single particle in the double-slit setup is also discussed. In
section V we draw our concluding remarks.

\section{Double-slit experiment modeled by initially correlated Gaussian wavepacket}

Consider a classical double-slit experiment with initially
correlated gaussian wavepacket. The initial $\sigma_{xp}$
correlation will be represented by the real parameter $\rho$ which
can take values in the interval $-\infty<\rho<\infty$
\cite{Dodonov}. Assume that such coherent correlated Gaussian
wavepacket of initial transverse width $\sigma_{0}$ is produced in
the source $S$ and propagates during a time $t$ before arriving at a
double-slit which splits it into two Gaussian wavepackets. After
crossing the grid the wavepackets propagate during a time $\tau$
before arriving at detector $D$ in detection screen, where they are
recombined. In this model we realize quantum effects only in
$x$-direction as we can assume that the energy associated with the
momentum of the particles in the $z$-direction is very high such
that the momentum component $p_{z}$ is sharply defined, i.e.,
$\Delta p_{z}\ll p_{z}$. Then we can consider a classical movement
in this direction at velocity $v_{z}$ and we may write $z=v_{z}t$
\cite{Paz3}. The sketch of this model is presented in Fig. 1.

\begin{figure}[htp]
\centering
\includegraphics[width=5.0 cm]{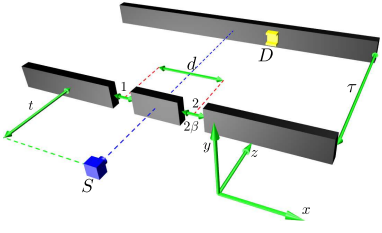}
\caption{Sketch of the double-slit experiment. The source $S$
produces a correlated Gaussian wavepacket with transverse width
$\sigma_{0}$. The wavepacket propagates during a time $t$ before
attaining the double-slit and during a time $\tau$ from the
double-slit to the detector $D$ in the screen of detection. The slit
transmission functions are taken to be Gaussian of width $\beta$ and
separated by a distance $d$.}\label{Figure1}
\end{figure}

The wavefunctions at the left $1(+)$ and right slit $2(-)$ are given
by \cite{Carol}

\begin{eqnarray}
\psi(x,t,\tau)&=&\int_{-\infty}^{\infty} dx_j\int_{-\infty}^{\infty}
dx_iG_2(x,t+\tau;x_j,t)\nonumber\\
&\times&F(x_j\pm d/2)G_1(x_j,t;x_i,0)\psi_0(x_i) , \label{M1}
\end{eqnarray}
 where

\begin{equation}
G_1(x_j,t;x_i,0)= \sqrt{\frac{m}{2\pi i \hbar t}} \exp
\left[\frac{im(x_j-x_i)^2}{2 \hbar t} \right]  ,  \label{M1}
\end{equation}

\begin{equation}
G_2(x,t+ \tau;x_j,t)= \sqrt{\frac{m}{2\pi i \hbar \tau}} \exp
\left[\frac{im(x-x_j)^2}{2 \hbar \tau} \right]  ,  \label{M1}
\end{equation}

\begin{equation}
F(x_j \pm d/2)= \frac{1}{\sqrt{\beta \sqrt{\pi}}}  \exp
\left[-\frac{im(x_j \pm d/2)^2}{2 {\beta}^2} \right]  ,  \label{M1}
\end{equation}

and
\begin{equation}
\psi_0(x_i)= \frac{1}{\sqrt{\sigma \sqrt{\pi}}}  \exp
\left[-\frac{{x^2_i}}{{2\sigma^2_0}} + \frac{i \rho
x^2_i}{2\sigma^2_0}\right]. \label{M1}
\end{equation}

The kernels $G_{1}(x_{j},t;x_{i},0)$ and $G_{2}(x,t+\tau;x_{j},t)$
are the free nonrelativistic propagators for a particle of mass $m$,
$F(x_{j}\pm d/2)$ describes the double-slit transmission functions
which are taken to be Gaussians of width $\beta$ separated by a
distance $d$. The parameter $\rho$ ensures that the initial state is
correlated. In fact, we obtain for the initial state $\psi_0(x_i)$
that the uncertainty in position is
$\sigma_{xx}=\sigma_{0}/\sqrt{2}$, whereas the uncertainty in
momentum is
$\sigma_{pp}=(\sqrt{1+\rho^{2}})\hbar/\sqrt{2}\sigma_{0}$ and the
$\sigma_{xp}$ correlations is $\sigma_{xp}=\hbar \rho/2$. For
$\rho=0$ we have a simple uncorrelated Gaussian wavepacket with
$\sigma_{xp}=0$. In order to obtain analytic expressions for the
wavefunction, Wigner function and  $\sigma_{xp}$ correlations in the
screen of detection we use a Gaussian transmission function instead
of a top-hat transmission one, because both a Gaussian transmission
function represents a good approximation to the experimental reality
and it is mathematically simpler to treat than a top-hat
transmission function.

After some algebraic manipulations, we obtain the following result
for the wavefunction that passed through slit $1(+)$
\begin{eqnarray}
\psi_1(x,t,\tau)&=&\frac{1}{\sqrt{B \sqrt{\pi}}}  \exp
\left[-\frac{{(x+D/2)}^2}{{2 B^2}} \right]\nonumber\\
&\times&\exp \left(\frac{i m x^2}{2 \hbar R} + i \Delta x + i \theta
+ i \mu \right) , \label{M1}
\end{eqnarray}

where

\begin{equation}
R (t,\tau)= \tau \frac{ \left(\frac{1}{\beta^2} + \frac{1}{b^2}
\right)^2 + \frac{m^2}{\hbar^2} \left(\frac{1}{ \tau} +
\frac{1}{r}\right)^2}{ \frac{1}{\beta^4} + \frac{C}{\sigma^4_0 (t^2
+ \tau^2_0 + 2 \tau_0 t \rho + t^2 \rho^2)}}, \label{M1}
\end{equation}

\begin{equation}
C=\left[ \tau^2_0 + \frac{t \tau^2_0}{\tau} + \tau^2_0 \rho^2 +
\frac{\tau^3_0 \rho}{\tau} + \frac{t \tau^2_0  \rho^2}{\tau} +
\frac{2\tau^2_0 \sigma^2_0}{\beta} \right],
\end{equation}

\begin{equation}
B^2(t,\tau)= \frac{ \left(\frac{1}{\beta^2} + \frac{1}{b^2}
\right)^2 + \frac{m^2}{\hbar^2} \left(\frac{1}{ \tau} +
\frac{1}{r}\right)^2}{ (\frac{m}{\hbar \tau})^2
\left(\frac{1}{\beta^2} + \frac{1}{b^2}  \right)}, \label{M1}
\end{equation}

\begin{equation}
\Delta(t,\tau)=\frac{\tau \sigma^2_0 d}{2 \tau_0 \beta^2 B^2},
\label{M1}
\end{equation}

\begin{equation}
D(t,\tau)=d \frac{\left(1+{\frac{\tau}{r}}\right)}{\left( 1+
\frac{\beta^2}{b^2}\right)}, \label{M1}
\end{equation}

\begin{equation}
\theta(t,\tau)= \frac{m d^2 \left( \frac{1}{\tau} +
\frac{1}{r}\right)}{ 8 \hbar \beta^4 \left[ \left( \frac{1
}{\beta^2} + \frac{1}{b^2}\right)^2 + \frac{m^2}{\hbar^2} \left(
\frac{1}{\tau} + \frac{1}{r}\right)^2 \right ] }, \label{M1}
\end{equation}

\begin{equation}
\mu(t,\tau)= - \frac{1}{2} \arctan \left[ \frac{ t + \tau \left( 1 +
\frac{\sigma^2_0}{\beta^2} + \frac{t \hbar \rho }{ m \beta^2 }
\right) } { \tau_0 \left( 1 - \frac{t \tau \sigma^2_0}{\tau_0
\beta^2} \right) + \rho \left( t + \tau \right)} \right], \label{M1}
\end{equation}

\begin{equation}
b(t)= \frac{\sigma_0}{\tau_0} \left[{ t^2 + \tau^2_0 + 2 t \tau_0
\rho + t^2 \rho^2 } \right]^\frac{1}{2}, \label{M1}
\end{equation}

and
\begin{equation}
r(t)= \frac{ \left(  t^2 + \tau^2_0 + 2 t \tau_0 \rho + t^2 \rho^2
\right) } { \left[  t \left( 1 +  \rho^2 \right) + \rho \tau_0
\right] } . \label{M1}
\end{equation}

For the right slit $2(-)$, we have just to substitute the parameter
$d$ with $-d$ in the expressions corresponding to the wave passing
through the first slit. Here, the parameter $B(t,\tau)$ is the beam
width for the propagation through one slit, $R(t,\tau)$ is the
radius of curvature of the wavefronts for the propagation through
one slit, $b(t)$ is the beam width for the free propagation and
$r(t)$ is the radius of curvature of the wavefronts for the free
propagation. $D(t,\tau)$ is the separation between the wavepackets
produced in the double-slit. $\Delta(t,\tau)x$ is a phase which
varies linearly with the transverse coordinate. $\theta(t,\tau)$ and
$\mu(t,\tau)$ are the time dependent phases and they are relevant
only if the slits have different widths. $\mu(t,\tau)$ is the Gouy
phase for the propagation through one slit. Differently from the
results obtained in Ref. \cite{solano}, all the parameters above are
affected by the correlation parameter $\rho$.
$\tau_{0}=m\sigma_{0}^{2}/\hbar$ is one intrinsic time scale which
essentially corresponds to the time during which a distance of the
order of the wavepacket extension is traversed with a speed
corresponding to the dispersion in velocity. It is viewed as a
characteristic time for the ``aging" of the initial state
\cite{Carol,solano} since it is a time from which the evolved state
acquires properties completely different from the initial state.

%%%%%%%%%%%%%%%%%%%%%%%%%%%%%%%%%%%%%%%%%%%%%%%%%%%%
\section{Wigner function and $\sigma_{xp}$ correlations}

Having obtained the wavefunctions in the previous section, we
calculate the Wigner function at the detection screen. Then we use
the Wigner function to evaluate the $\sigma_{xp}$ correlations. We
observe that the $\sigma_{xp}$ correlations have a minimum and a
maximum as a function of the propagation time $t$ for a negative
value of the correlation parameter $\rho$. For positive or null
values of $\rho$ such correlations have only a maximum. We also
notice that the Wigner function for the state with minimum
$\sigma_{xp}$ correlations does not present a negative part and thus
it represents a classical state in this sense. On the other hand,
the Wigner function for the state with $\rho\geq 0$ presents a
negative part for some values of position $x$ and momentum $p$ and
thus signalize quantum state \cite{Kenfack}. The state with negative
correlation parameter $\rho<0$ is known as contractive state
\cite{Storey} which, for a free quantum particle, was introduced by
Yuen \cite{Yuen}  in an attempt to evade the standard quantum limit
for repeated position measurements.

The Wigner function of the phase-space quasiprobability distribution
 in one-dimension configuration space is
defined as \cite{Wigner}
\begin{eqnarray}
W(x,k) = \frac{1}{2 \pi} \int_{-\infty}^{\infty} d y  e^{-i k y}
\psi^* \left(  x-\frac{y}{2} \right) \psi \left( x + \frac{y}{2}
\right), \label{wigner}
\end{eqnarray}
where
\begin{equation}
\psi(x,t,\tau)= \frac{\psi_1 (x,t,\tau) + \psi_2 (x,t,\tau)}
{\sqrt{2 + 2 \exp \left[ -  \frac{D^2}{4 B^2}  - \Delta^2 B^2
\right]}}, \label{M1}
\end{equation}
is the normalized wavefunction at the screen of detection of the
double-slit experiment.

By solving the integration in equation (\ref{wigner}) we obtain the
following result
\begin{eqnarray}
W(x,k)&=&W_{1}(x,k)+W_{2}(x,k)\nonumber\\
&+&\frac{2}{\pi\alpha^{2}}\exp\left[-\left(\frac{x^2}{B^2}+\left(k-\frac{mx}{\hbar
R}\right)^2\right)B^2\right]\nonumber\\
&\times&\cos\left[\left(k-\frac{mx}{\hbar R}\right)D+2\Delta
x\right] \label{wigner1}
\end{eqnarray}
where
\begin{eqnarray}
W_{1}(x,k)&=&\frac{1}{\pi\alpha^{2}}\exp\left[-\frac{\left(x+\frac{D}{2}\right)^2}{B^2}\right]\nonumber\\
&\times&\exp\left[-\left(k-\frac{mx}{\hbar
R}-\Delta\right)^2B^2\right],
\end{eqnarray}

\begin{eqnarray}
W_{2}(x,k)&=&\frac{1}{\pi\alpha^{2}}\exp\left[-\frac{\left(x-\frac{D}{2}\right)^2}{B^2}\right]\nonumber\\
&\times&\exp\left[-\left(k-\frac{mx}{\hbar
R}+\Delta\right)^2B^2\right],
\end{eqnarray}
and
\begin{equation}
\alpha = 2 + 2 \exp \left[ - \frac{  D^2 }{ 4 B^2 } -  \Delta^2 B^2
\right].
\end{equation}
The result displayed by equation (\ref{wigner1}) is composed by the
terms $W_{1}(x,k)$ and $W_{2}(x,k)$ which are the Wigner functions
for a particle that passed through  slit $1$ and $2$, respectively,
as well as an interference term.

The Wigner function and the $\sigma_{xp}$ correlations are related
by \cite{Lerner}
\begin{equation}
\sigma_{xp}=\frac{\int xpW(x,p)dxdp}{\int W(x,p)dxdp},
\end{equation}
where $p=\hbar k$. After some algebraic manipulation we obtain
\begin{eqnarray}
\sigma_{xp}(t, \tau)&=& \frac{m B^2}{2 R} + \frac{ \left( m D^2/R
\right)}{4  + 4 \exp \left[- \frac{D^2}{4B^2} - \Delta^2 B^2
\right]} - \frac{\hbar \Delta D }{2}\nonumber\\
 &-& \frac{\left(   m
\Delta^2 B^4 /R \right)}{1+ \exp \left[ \frac{D^2}{4 B^2} + \Delta^2
B^2 \right]}.  \label{xp}
\end{eqnarray}

In the following, we plot the curve for the $\sigma_{xp}$
correlations as a function of the time $t/\tau_{0}$ for neutrons.
The reason to consider neutrons relies in their experimental
reality, which is most close to our model for interference with
completely coherent matter waves, although we still have loss of
coherence as discussed in Ref. \cite{Sanz}. We adopt the following
parameters: mass $m=1.67\times10^{-27}\;\mathrm{kg}$, initial width
of the packet $\sigma_{0}=7.8\;\mathrm{\mu m}$ (which corresponds to
the effective width of $2\sqrt{2}\sigma_{0}\approx22\;\mathrm{\mu
m}$), slit width $\beta=7.8\;\mathrm{\mu m}$, separation between the
slits $d=125\;\mathrm{\mu m}$ and de Broglie wavelength
$\lambda=2\;\mathrm{nm}$. These same parameters were used previously
in double-slit experiments with neutrons by A. Zeilinger and
collaborators. \cite{Zeilinger1}. In Fig. 2, we show the
correlations as a function of $t/\tau_{0}$ for $\tau=18\tau_{0}$ and
$\rho=-1.0$. We use a negative value of the correlation parameter
$\rho$ in order to obtain $\sigma_{xp}$ correlations with a point of
minimum and of maximum. For the parameters above we calculate the
points of minimum and maximum correlations and obtain, respectively,
$t_{min}=0.49\tau_{0}$ and $t_{max}=1.36\tau_{0}$.

\begin{figure}[htp]
\centering
\includegraphics[width=5.0 cm]{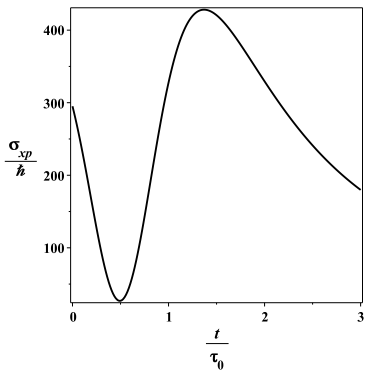}
\caption{$\sigma_{xp}$ correlations as a function of $t/\tau_{0}$
for $\tau=18\tau_{0}$ and $\rho=-1.0$.}
\end{figure}

In Fig. 3 we show the Wigner function as a function of $x$ and $k$
for the parameters which characterize neutrons. In Fig. 3(a) we have the Wigner
function for the time for which the $\sigma_{xp}$ correlations are
minima and in Fig. 3(b) we show the Wigner function at the time for
which these correlations are maxima. We observe that (maximum) minimum
$\sigma_{xp}$ correlations is associated with a (negative) positive definite
Wigner function. In fact, a positive Wigner
function is obtained only for $t_{min}$, the time for which the
$\sigma_{xp}$ correlations is minimum, and for any time different of
$t_{min}$ the Wigner function has a negative portion. We also observe
that the maximum negative portion occurs for $t_{max}$ the time for
which the $\sigma_{xp}$ correlations is maximum.

\begin{figure}[htp]
\centering
\includegraphics[width=4.2 cm]{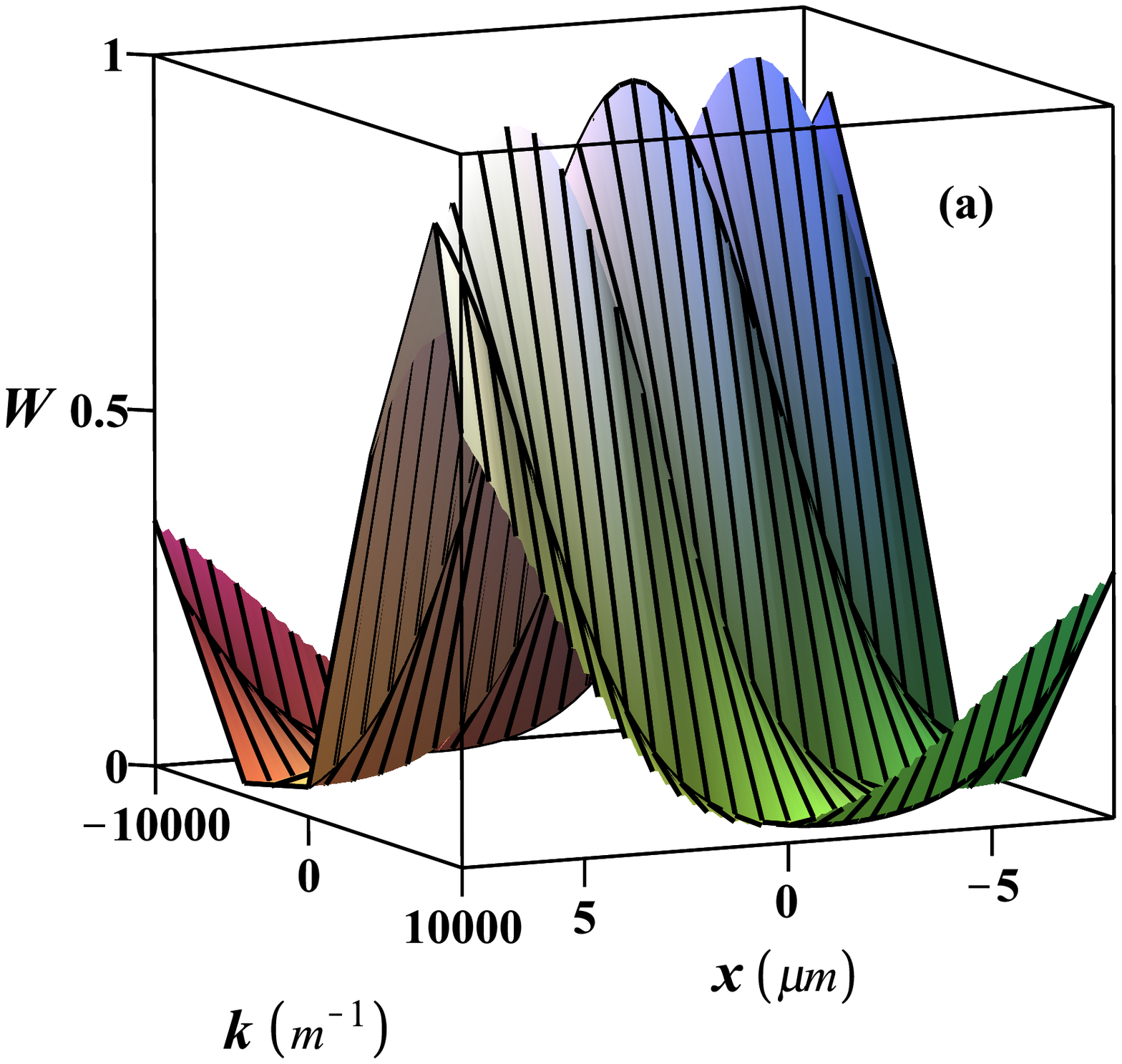}
\includegraphics[width=4.2 cm]{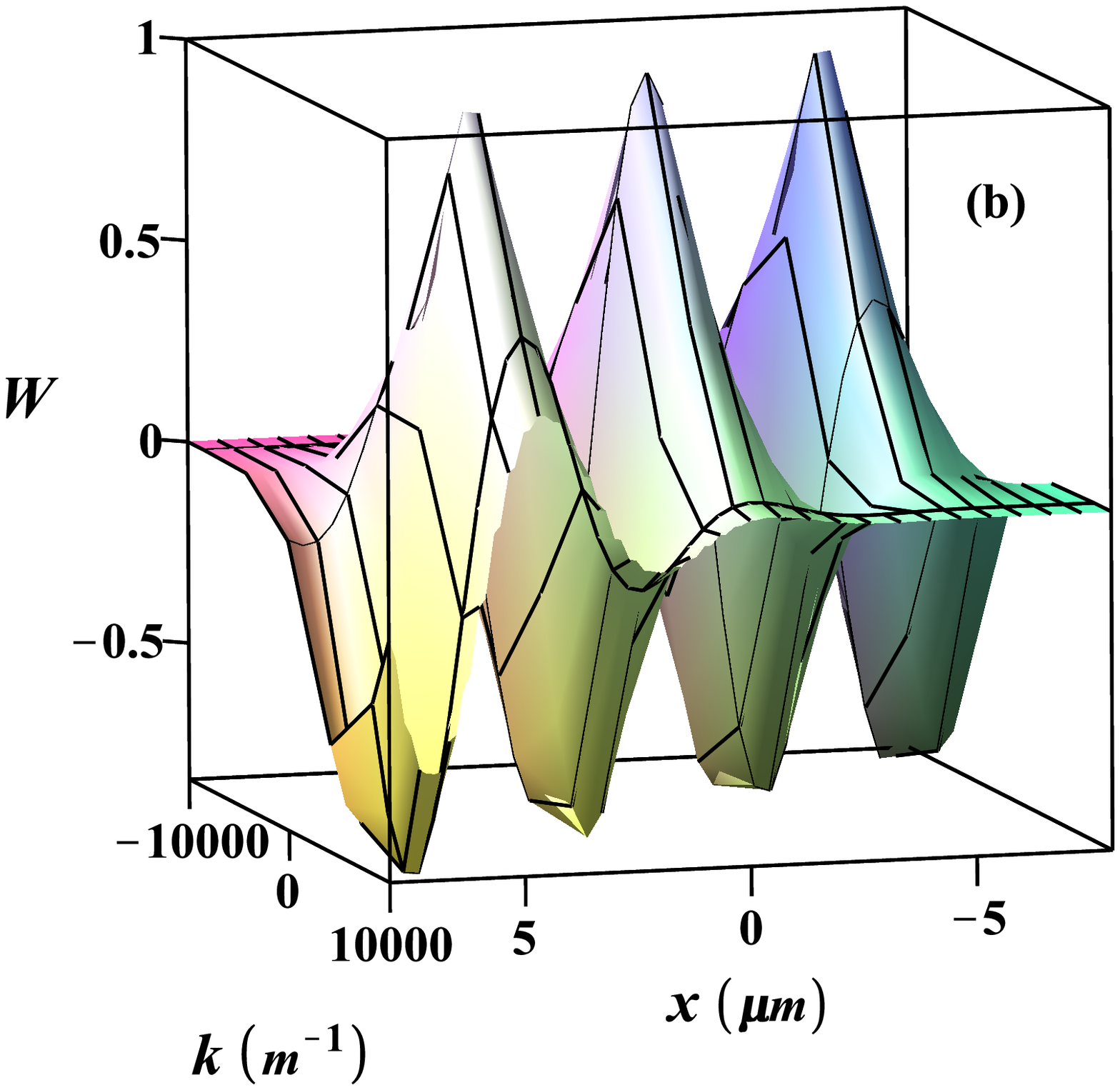}
\caption{Wigner function as a function of $x$ and $k$. (a) Wigner
function for the time for which the $\sigma_{xp}$ correlations are
minima and (b) Wigner function for the time for which these
correlations are maxima.}
\end{figure}

Therefore a single particle in the double-slit setup can present
Wigner function distribution at the screen of detection with
different behaviors, depending on the initial $\sigma_{xp}$
correlations as well as of its dynamics. The correlations at the
times of minimum and maximum are governed, respectively, by the
first and second terms of equation (\ref{xp}), i.e.,
$\sigma_{xp}(t_{min})\approx m B^2/2R$ and
$\sigma_{xp}(t_{max})\approx m D^2/4R$. Therefore, we have the
following conditions
\begin{equation}
B^2(t_{min})\gg D^2(t_{min}),\;\; D^2(t_{max})\gg B^2(t_{max}).
\label{cond}
\end{equation}
Since $B(t,\tau)$ is the width of the wavepacket and $D(t,\tau)$ the
separation between the wavepackets at the screen, the region of
overlap between the two packets is bigger for  minimum $\sigma_{xp}$
correlations as compared to maximum correlations. When we apply the
conditions of equation (\ref{cond}) in the terms of equation
(\ref{wigner1}) we observe that the interference term contribute to
the Wigner function more than the terms $W_{1}(x,k,t)$ and
$W_{2}(x,k,t)$.

The state with minimum $\sigma_{xp}$ correlations is a superposition
of two Gaussian wavefunctions but has a Gaussian shaped Wigner
function exhibiting only positive values. This is analogous of the
original EPR state for a single particle. Most interesting here is
verify whether these states violate a Bell inequality enabling us to
demonstrate the nonlocal character for a single particle
correlation, just as for the analogous original EPR state, with
possibility of measurement in the double-slit setup. In order to
show nonlocal character for a single particle we construct in the
next section its Bell's inequality.

\section{Nonlocal correlations and Bell inequality}

Since Bell type inequalities can be found for the correlations of
spin and polarization of a single particle the same should be true
for the correlations of other observables such as position $\hat{x}$
and momentum $\hat{p}$ operators as in the original EPR paper
\cite{EPR}. In order to attest the nonlocal correlations for these
observables, in this section we construct the Bell's inequality using
the Wigner function of the wavefunction at the screen of detection
in the double-slit experiment. We obtain the Bell's inequality
violation for both positive and negative Wigner function which
characterizes nonlocal correlations of position and momentum
observables of a single particle. We define the correlation function
as in Ref. \cite{Aspect}. We consider the following four
combinations of $x$ and $k$: $x_{1}=x_{3}$, $x_{2}=x_{4}=x$,
$k_{1}=k_{2}$ and $k_{3}=k_{4}=k$. With these quantities we
construct the combination
\begin{equation}
\mathcal{B}(t,x,k)=|W(x_{1},k_{1})+W(x,k_{1})|+|W(x_{1},k)-W(x,k)|,
\label{Bell}
\end{equation}
which was pointed out as similar to the CHSH inequality
\cite{Anderson}.

We observe that the terms $W_{1}(x,k)$ and $W_{2}(x,k)$ in equation
(\ref{wigner1}) can be neglected in comparison with the interference
term. Therefore, the interference causes the Bell's inequality
violation in equation (\ref{Bell}) namely a nonlocal correlation. In \cite{Isobe} it
was shown that the BCHSH inequality violation stems from an
interference effect as well.

In Fig. 4 we show the Bell inequality as a function of $x$ and $k$
for the parameters of neutrons with $x_{1}=1\;\mathrm{\mu m}$ and
$k_{1}=10^{3}\;\mathrm{m^{-1}}$. In Fig. 4(a) we have the Bell
inequality at the time for which the $\sigma_{xp}$ correlations are
minimal and in Fig. 4(b) we show the Bell inequality for the time
for which these correlations are maximal. We show only values
exceeding the bound imposed by local theories. We observe that the
state for $t_{max}$ violates the Bell inequality more than the state
for $t_{min}$ for specific values of $x$ and $k$. We also observe
that the maximum value of the Bell inequality for the state when the
time is $t_{min}$, the original EPR state, is
$\mathcal{B}\approx2.19$. The same value was obtained in Ref.
\cite{Banaszek} using another system.

\begin{figure}[htp]
\centering
\includegraphics[width=4.2 cm]{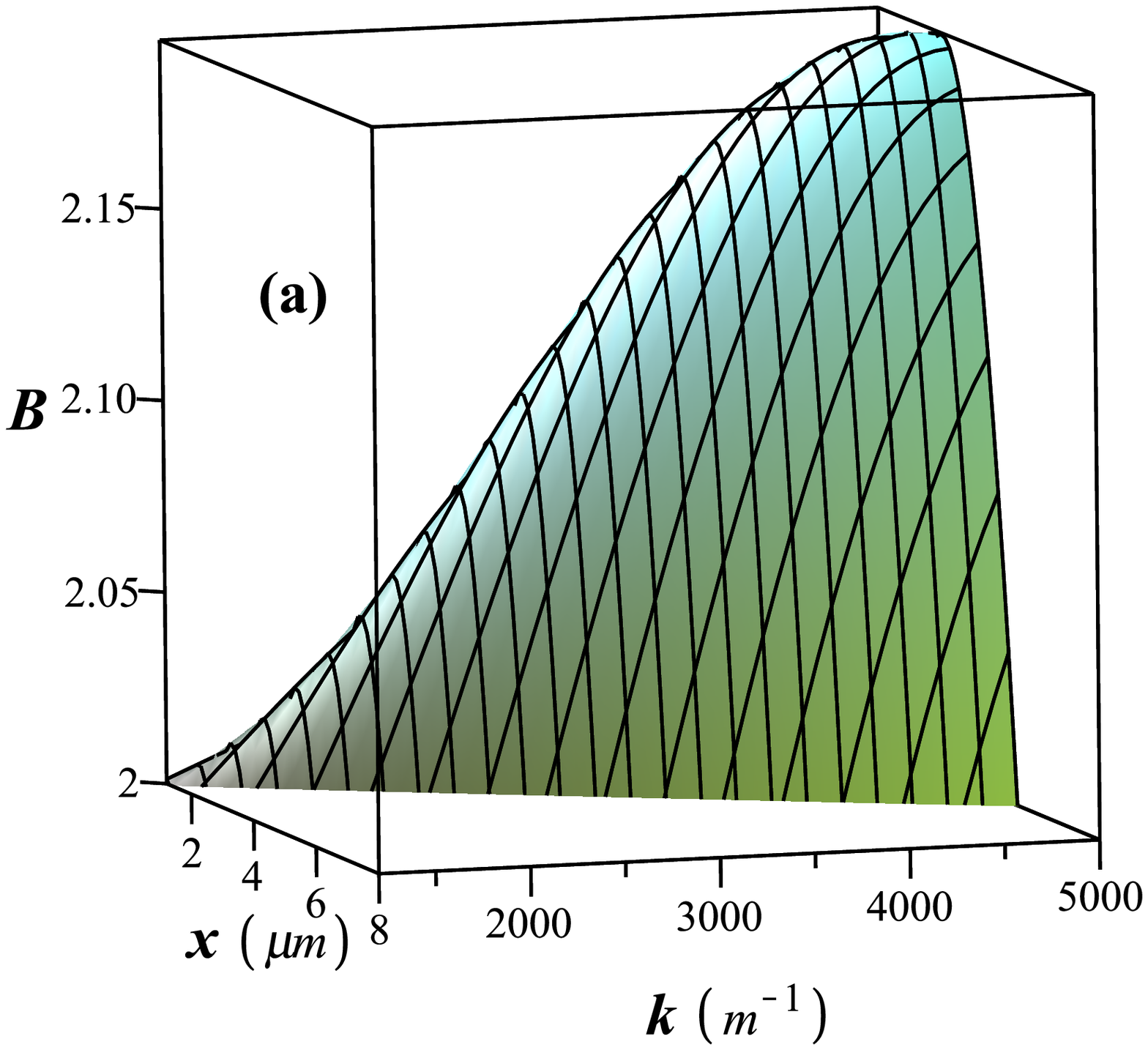}
\includegraphics[width=4.2 cm]{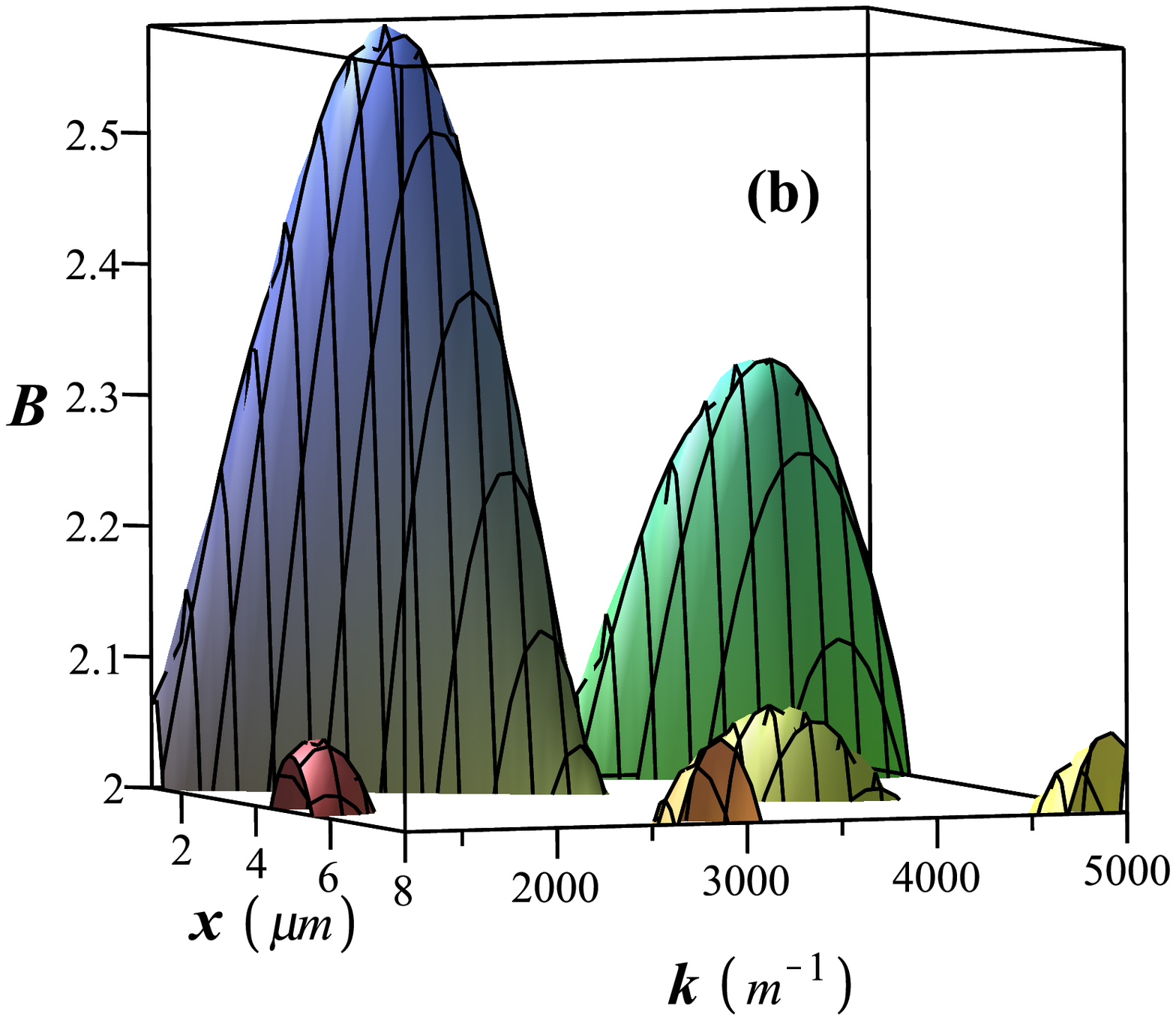}
\caption{Bell's inequality as a function of $x$ and $k$. (a) Bell's
inequality for $t_{min}$. The maximum violation is
$\mathcal{B}=2.19$. (b) Bell's inequality for $t_{max}$. The maximum
violation is $\mathcal{B}=2.59$ We show only values exceeding the
bound imposed by local theories.}
\end{figure}
Therefore, these results show the nonlocal character of correlations
for a single particle and enable us to define an analogous original
EPR state for it. Both positive and negative parts of the Wigner
function violate the Bell's inequality yet the negativity
contributes more for the Bell's inequality violation. The maximum
violation for the minimum of $\sigma_{xp}$ correlations is
$\mathcal{B}=2.19$ and for maximum $\sigma_{xp}$ correlations is
$\mathcal{B}=2.59$. The Bell's inequality is a result of the
interference in the double-slit experiment which is a consequence of
wavy property. As the double-slit experiment is simple to implement
and  the Wigner function can be experimentally accessed, the results
above can in principle be experimentally tested. Thus, these results
show nonlocality of noncommuting observables associated with a
single particle just as the Bell's inequality violation for the
original EPR two-particle state. The analogy of two particles EPR
state was also found previously by considering the atom-field
interaction, although the Bell's inequality violation was not
constructed for this system \cite{Storey}.

\section{Conclusions}
We showed that the $\sigma_{xp}$ correlations at the screen of
detection in the double-slit experiment can be maxima and minima
depending of the time evolution from the source to the double-slit.
We obtained that minimal correlations are possible only if the
initial state is a contractive state, i.e., the state with negative
coefficient of correlation. We observed that there is a connection
between the $\sigma_{xp}$ correlations and the positivity/negativity
of the Wigner function. The minimal correlations are associated with
a positive definite Wigner function whereas correlations other than
those are associated with negative parts in the Wigner function. The
maximal correlations are associated with maximal negative parts in
the Wigner function. We used the Wigner function to construct a
Bell-type inequality for the noncommuting position and momentum
observables. The Bell's inequality is violated for positive and
negative Wigner function for some values of the position and
momentum variables. Therefore, we have shown the nonlocal character
of a single particle in the double-slit experiment. By choosing a
certain set  of parameters the maximum Bell's inequality violation
is $\mathcal{B}=2.59$ and is obtained when the Wigner function has
greater negative part. On the other hand, when the Wigner function
is positive definite the maximum Bell's inequality violation is
$\mathcal{B}=2.19$. The case of positive Wigner function is the
analogous of the original EPR state for a single particle.
Therefore, our results display nonlocality of the position and
momentum observables for a single particle which can be tested in
the double-slit experiment.

\vskip1.0cm
\begin{acknowledgments}
O. R. de Araujo thanks the CAPES by financial support under grant
number 210010114016P3. H. Alexander and Marcos Sampaio thank the
CNPq by financial support. 
\end{acknowledgments}

%\pagebreak

\end{document}